\documentstyle[12pt, epsf]{article}
\input math_macros.tex

\def\ref#1{$^{#1)}$}

\begin{document}
\begin{titlepage}
\begin{center}
\today     \hfill    LBL-35972 \\

\vskip .15in

{\large \bf Quantum
 Electrodynamics at Large Distances II: Nature of the Dominant Singularities.}
\footnote{This work was supported by the Director, Office of Energy
Research, Office of High Energy and Nuclear Physics, Division of High
Energy Physics of the U.S. Department of Energy under Contract
DE-AC03-76SF00098, and by the Japanese Ministry of Education,
 Science and Culture under a
Grant-in-Aid for Scientific Research
(International Scientific Research Program 03044078).}

\vskip .15in
Takahiro Kawai\\

{\em Research Institute for Mathematical Sciences\\
Kyoto University\\
Kyoto 606-01 JAPAN}

\vskip .15in

Henry P. Stapp \\

{\em Lawrence Berkeley Laboratory\\
     University of California\\
     Berkeley, California 94720}

\end{center}

\vskip .15in

\begin{abstract}

Accurate calculations of macroscopic and mesoscopic properties
in quantum electrodynamics require  careful treatment of infrared
divergences: standard treatments introduce spurious
large-distances effects. A method for computing these properties was
developed in a companion paper. That method depends upon a result obtained
here about the nature of the singularities that produce the dominant
large-distance behaviour. If all particles in a quantum field theory have
non-zero mass then the Landau-Nakanishi diagrams give strong conditions on the
singularities of the scattering functions. These conditions are severely
weakened in quantum electrodynamics by effects of points where photon
momenta vanish. A new kind of Landau-Nakanishi diagram is
developed here. It is geared specifically to the pole-decomposition functions
that dominate the macroscopic behaviour in quantum electrodynamics, and
leads to strong results for these functions at points where photon momenta
vanish.

\end{abstract}
\end{titlepage}
\renewcommand{\thepage}{\roman{page}}
\setcounter{page}{2}
\mbox{ }

\vskip 1in

\begin{center}
{\bf Disclaimer}
\end{center}

\vskip .2in

\begin{scriptsize}
\begin{quotation}
This document was prepared as an account for work sponsored by the United
States Government.  Neither the United States Government nor any agency
thereof, nor The Regents of the University of California, nor any of their
employees, makes any warranty, express or implied, or assumes any legal
liability or responsibility for the accuracy, completeness, or usefulness
of any information, apparatus, product, or process disclosed, or represents
that its use would not infringe privately owned rights.  Reference herein
to any specific commercial products process, or service by its trade name,
trademark, manufacturer, or otherwise, does not necessarily constitute or
imply its endorsement, recommendation, or favoring by the United States
Government or any agency thereof, or The Regents of the University of
California.  The views and opinions of authors expressed herein do not
necessarily state or reflect those of the United States Government or any
agency thereof of The Regents of the University of California and shall
not be used for advertising or product endorsement purposes.
\end{quotation}
\end{scriptsize}

\vskip 2in

\begin{center}
\begin{small}
{\it Lawrence Berkeley Laboratory is an equal opportunity employer.}
\end{small}
\end{center}

\newpage
\renewcommand{\thepage}{\arabic{page}}
\setcounter{page}{1}
\noindent

\noindent {\bf 1. Introduction}

A method of calculating the macroscopic and mesoscopic properties of
scattering functions in quantum electrodynamics was developed in reference 1,
in the context of a particular example. The large-distance  behaviour
was shown to be concordant with the idea that electrons propagate over large
distance like  stable particles in classical physics. This result is expected,
and indeed is required in the interpretation of scattering experiments. But
unless one is able to {\it deduce} this dominant behaviour from the theory,
and exhibit a controlled non-dominant remainder, the theory would be
unsatisfactory, for it
would lack the power to make valid predictions in the mesoscopic regime
lying between the quantum and classical realms. This regime is becoming
increasingly important for technology.

The extraction from quantum electrodynamics of the correspondence-principle
large-distance part plus a well-controlled non-dominant remainder is a not a
trivial
exercise.  Difficulties arise from: 1), the spurious large-distance effects
introduced by the usual momentum-space treatments of infrared
divergences; 2), the singular character of the photon-propagator singularity
surface $k^2=0$ at $k=0$; 3), the occurrence of several different
types of singularities on certain singularity surfaces; and 4), the need to
deal
effectively with the pole-decomposition functions that control the
large-distance properties. These problems were all
dealt with in reference 1. But one key property was left unproved. The
immediate aim of this paper is to establish this property. In the course
of doing so we shall develop powerful methods for dealing with singularities
arising in quantum electrodynamics.

A first problem to be faced is the weakening of the Landau-Nakanishi
diagrammatic conditions for the presence of a singularity. The vanishing of the
gradient of $k^2$ at $k=0$ renders the original versions$^{3,4}$ of these
conditions trivial: they yield no condition at all, for functions
that describe processes with internal photons. Improved versions that cover the
$k=0$ points have been devised$^{5}$. But these also have too many solutions:
in general a {\it continuum} of essentially different diagrams all lead to any
given point on the Landau singularity surface. This surplus of diagrams
precludes the application of the simple known rule$^{6}$ for the
nature of the singularity on that surface.

The first part of our resolution  of the problem is this: Use not the original
momentum-space variables, but rather a set of nested radial coordinates
and the associated angles. These variables are defined by first separating the
integration
region into sectors specified by the different orderings of the relative sizes
of the Euclidean norms $|k_i|$ of the soft-photon energy-momenta $k_i$;
then, in each sector, re-ordering the vectors $k_i$ by size, so that
$|k_i|\geq |k_{i+1}|$; and
finally writing
$$
k_i= r_1r_2...r_i\Omega_i,\eqno(0)
$$
where, for all $i$, $|\Omega_i|=1$ and $0\leq r_i \leq 1$.

A second problem is that we need results not for the scattering
functions themselves but rather to the functions obtained from them by
decomposing their meromorphic parts into sums of poles times residues. The
functions obtained by this pole decomposition  give the dominant
large-distance behaviour. We devise a new kind of ``Landau'' diagram
for these functions.

The specific example considered in reference 1 pertains to a Feynman
graph consisting of six hard photons coupled at six vertices
into a single charged-particle closed loop. These six vertices
are divided into three disjoint pairs, with the two vertices in each pair
linked by a charged-particle
line that is associated with a momentum-energy vector that is
far off mass shell. This line can, for our purposes, be
shrunk to a point. This produces a (triangle) graph $G$ consisting of three
internal charged-particle lines, with two hard photons attached at each of the
three vertices.

We now ``dress'' this triangle graph $G$ with soft photons: we consider the
set of graphs $\{g\}$
obtained by coupling all possible numbers of soft photons into this
charged-particle loop in all possible ways. If we separate the interaction
term $ie\gamma_{\mu}$ into its classical and quantum parts, in the way
described in ref. 1, then all the classical
interactions can be shifted to the three hard vertices, leaving only quantum
vertices along the three sides of the triangle. Each of these three sides $s$
of
the original triangle graph $G$ is therefore now divided into segments
by a set of quantum vertices. Each segment $j$ is associated with a Feynman
denominator $(p_s+K_j)^2-m^2+i0$, where $K_j$ is some (algebraic) sum of photon
momenta. The total contribution from all ``classical photons'', which are the
photons that are coupled into $G$ {\it only} at classical vertices, can be
factored off as a single unitary operator that is independent of the
non-classical remainder.

We are interested here  in the properties of the individual terms of the
perturbation expansion of this remainder. Each such term is represented by
a Feynman graph $g$. Each soft photon is coupled on one or both ends into
either a vertex or a side of the original triangle graph $G$, with $C$
couplings at vertices and $Q$ couplings on the sides.

To exhibit what are expected to be (and turn out to be) the dominant
contributions to the singularity of the scattering function on the
triangle-diagram singularity surface $\varphi=0$ we consider the Feynman
denominator associated with each segments $j$ of side $s$ to be a pole in the
$z_s=p_s^2$ plane, and then express the function associated with each of the
three sides $s$ of the triangle as a sum over pole contributions:
$$
N_s \prod_{j=0}^n [(p_s+K_j)^2-m^2+i0]^{-1}=N_s \sum_{i=0}^n
[((p_s+K_i)^2-m^2+i0)D_{si}]^{-1}, \eqno(0')
$$
where $D_{si}$ is the product over $j\neq i$ of factors
$((p_s+K_j)^2-(p_s+K_i)^2)$.

There is a pole-decomposition formula like this for each of the three sides $s$
of the triangle. The direct aim of this paper is to show that for each
term consisting of photon propagators, together with three factors
$f_{s,i(s)}$, one from each side $s$ of $G$, with
$f_{s,i(s)}$ being the $i(s)$th term in the pole-decomposition formula $(0')$
associated with side $s$, the contours in $\Omega_i$-space can be shifted so
as to avoid, simultaneously, all singularities in the photon propagators
and residue factors. This result plays a crucial role in our arguments. It
means, for the case under study, that the part of the scattering function that
comes from the meromorphic parts of the propagators can be expressed as
a sum of terms, in each of which the only singularities are end-point
singularities at $r_i=0$ and $r_i=1$, and three Feynman denominators, one
for each of the three sides $s$ of the triangle $G$. The problems are
thereby focussed on the effects of the integrals over the $r_i$. These are
the issues resolved in papers I and III.
\newpage
\noindent {\bf 2. Notation}

\begin{figure}
\caption{
  The basic charged--particle triangle graph $G$.
The momentum--energy $p_s$ flows along side $s$ of the triangle in the
direction of the arrow.
The three energy components satisfy $p^0_1 > 0,$  $p_2^0<0$, and  $p^0_3>0$.}
\epsfxsize = 6.35in
\epsfysize = 3.53in
\epsffile{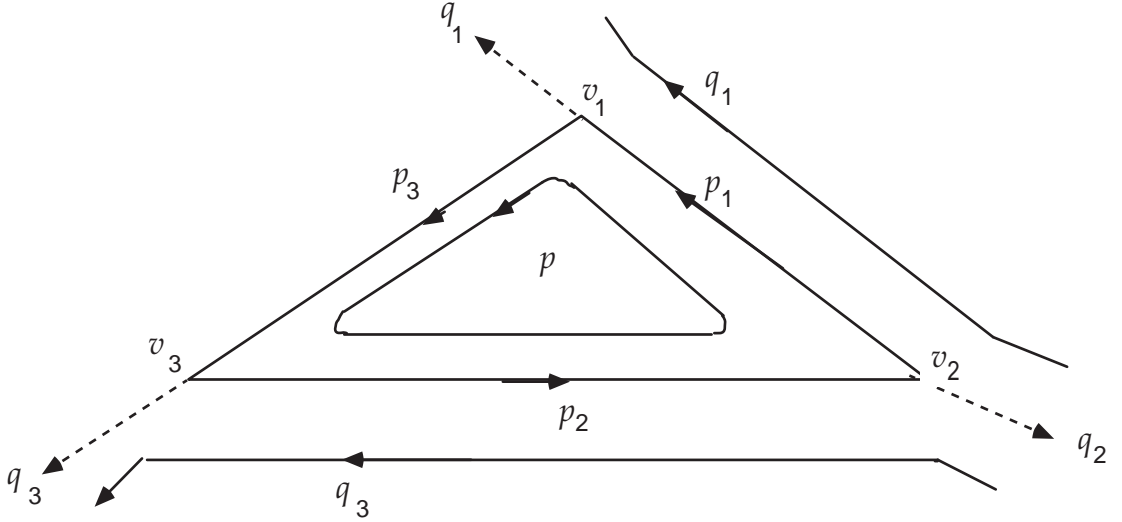}
\label{fig1}
\end{figure}

The original triangle graph $G$ is shown in \ref{fig1}.

The momenta $p_1, - p_2$, and $p_3$ represent the momenta flowing from $v_2$ to
$v_1$, from $v_2$ to $v_3$, and from $v_1$ to $v_3$,
respectively.
Conservation of energy--momentum is represented by introducing a closed loop
carrying momentum $p$, and two open paths carrying momenta $q_1$ and $q_3$,
respectively, in the directions indicated by the arrows.
Then $p_1 = p+q_1, \ \ \ p_2 = p - q_3$, and $p_3 = p$.

The function associated with this Feynman graph $G$ has a singularity on the
positive--$\alpha$ Landau--Nakanishi triangle--diagram singularity  surface
$\varphi (q) = 0,$
where $q=(q_1, q_2, q_3) $ and $q_3\equiv - q_1 - q_2$.
For each point $q$ on this surface $\varphi =0$ there is$^2$ a uniquely
defined set
of three four--vectors $p_1(q), p_2(q)$, and $p_3(q)$ such that the singularity
at $q$ of the Feynman function $F(G)$ corresponding to the graph $G$ of
Fig. 1 arises from an arbitrarily small neighborhood
$$
p \approx p(q)=p_1(q)-q_1=p_2(q)+q_3 = p_3(q)\eqno(1a)
$$
in the domain of integration of the Feynman function.
These three four--vectors $p_s(q)$ satisfy the mass--shell constraints
$$
(p_s(q))^2=m^2,\eqno(1b)
$$
and the (Landau--Nakanishi) loop equation
$$
\alpha_1 p_1(q)+\alpha_2p_2(q)+\alpha_3 p_3(q)=0,\eqno(1c)
$$
where the $\alpha_s$ are nonnegative real numbers.
This loop equation implies that for each $q$ on $\varphi (q)=0$ the three
four--vectors $p_s (q)$ lie in some two--dimensional subspace of the
four--dimensional energy--momentum space.

We shall consider a fixed {\it interior} point $q$ of the surface $\varphi =0$.
In this case each of the three parameters $\alpha_s$ is nonzero, and each of
the three four--vectors vectors $p_s(q)$ is nonparallel to each of the other
two.

Consider now a graph $g$ obtained by inserting some finite number of
soft--photon lines $i$ ($i\epsilon I$) into $G$.
Each inserted line begins on a line of $G$ and ends on a line of $G$.
The bound $\delta$ on the Euclidean norms $\abs{k_i}$ of the (soft)
 photon momenta is taken small enough so that
$$
n\delta < \delta^\prime  << m,\eqno(2)
$$
where $n$ is the number of photon lines in the graph.

The case under consideration here is one where every coupling is a $Q$-type
coupling. For a $C$-type coupling the corresponding vertex lies on one of the
three vertices of the graph $G$. The present argument can be carried over to
the case with some $C$-type couplings by simply contracting to points
some segments representing residue factors, thereby bringing each of various
vertices lying sides of $G$ into coincidence with a vertices of $G$. These
contractions (performed after the loops have been specified) do not upset the
arguments.

Momentum--energy conservation is now maintained by introducing a separate
closed loop for the momentum $k_i$ of each photon line.
Momentum $k_i$ flows along the photon line segment $i$ in the direction
indicated by the arrow placed on that line segment.
It then continues to flow through the graph $g$ by flowing along certain
charged-particle lines of this graph.
This continuation through $g$ is specified by the condition that this flow
line  pass through
at most one of the three vertices $v_1, v_2, v_3$.

The arrow on photon line $i$ is chosen so that every term $p_sk_i$ that occurs
in any Feynman denominator occurs with a plus sign.
Consequently, the Feynman rule that $m^2$ represents $m^2 -i0$ is compatible
with the rule that each $p_sk_i$ represents $p_sk_i + i0$.
No condition is placed on the sign of the energy component $k_i^0$.

Each charged--particle line segment $j$ has an arrow placed on it.
The momentum flowing along the charged--particle segment $j$ in the direction
of this arrow is called $\sum_j$.
It is the momentum $p_s$ flowing along the side of the triangle upon which
segment $j$ lies, as  defined in Fig. 1, plus the (algebraic) sum $K_j$
of the photon momenta $k_i$ carried by the photon
loops that pass along this segment $j$.

\begin{figure}
\epsfxsize = 4.51in
\epsfysize = 5.88in
\epsffile{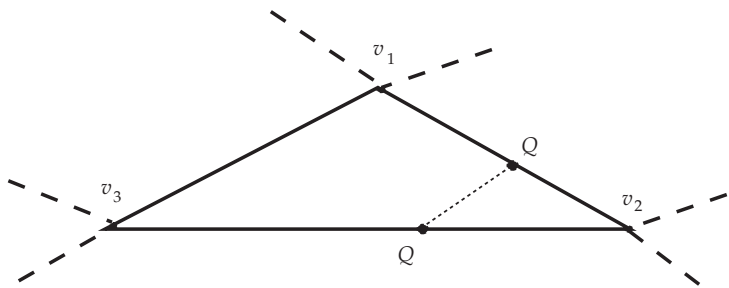}
\caption{ A graph $g$ representing a soft-photon correction to a hard-photon
triangle-diagram process $G$. Hard and soft photons are represented by dashed
and wiggly lines, repectively.}
\label{fig2}
\end{figure}

Our interest here is in the functions that arise from inserting the
pole-decomposition formula $(0')$ [or (5.5) of ref. 1] into the
meromorphic parts of the generalized propagators
corresponding to the three sides of the original triangle graph $G$.
Consider, for example, the simple graph $g$ of \ref{fig2}

\begin{figure}
\caption{ The $*$ graphs representing the four terms that arise from inserting
the pole-decomposition formula $(0')$ into the meromorphic part of the function
represented by the graph $g$ of Fig. 2.}
\epsfxsize = 6.35in
\epsfysize = 4.53in
\epsffile{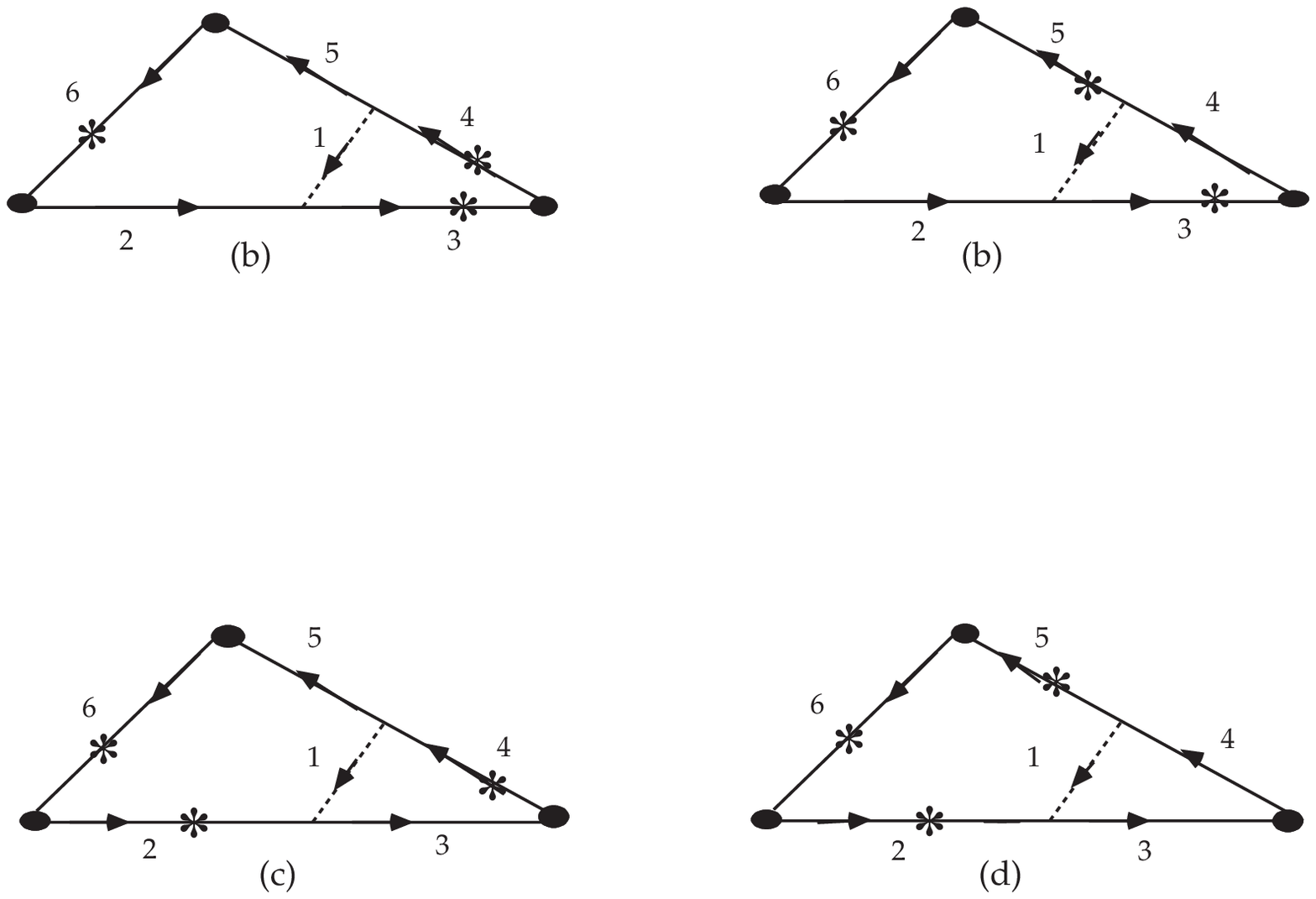}
\label{fig3}
\end{figure}

The meromorphic part of the function represented by the graph $g$ of Fig. 2 is
a sum of the four terms represented by the four $*$ graphs of \ref{fig3}

\noindent The asterisk $(*)$ on a line segment of a $*$ graph indicates that
it is the
segment associated with the (pole) denominator $(p_s+K_i)^2 -m^2+i0$ in the
pole-decomposition formula $(0')$.
Each of the other charge--particle segments $j\neq i$ is associated with a
pole--residue denominator function
$$
f_j =2 (p_s+K_i) \Omega_{ij} + \rho_{ij}\Omega^2_{ij} + i0 , \eqno(3a)
$$
where
$$
\rho_{ij}=r_1r_2... r_{\ell(i,j)}\eqno(3b)
$$
and $$
\Omega_{ij} = (\Omega_{\ell(i,j)}+...)= \sigma_{ij} (K_j-K_i)/\rho_{ij}.
\eqno(3c)
$$
The index $\ell(i,j)$ is the smallest $j$ such that $k_j$ appears in $K_i$
or  $K_j$, but not both.
Each of the non-exhibited terms in the parentheses in (3c) is a product of
some $\pm\Omega_k$ with a product of a non-empty set of factors $r_h (h\geq
2)$.

Each of the pole-residue factors $f_j$ is
formed by first taking the difference
$\sigma_{ij} (\sum^2_j-\sum^2_i)$, where $\sum_j=p_s+K_j$ is the
momentum--energy flowing along segment $j$ in the direction of the arrow on
that segment, and $\sum_i=p_s+K_i$ is the
momentum--energy flowing along the $*$ segment on the same side $s$ of the
charged--particle triangle, and then dividing out the common factors $r_h (h
\geq 1$). The sign
$\sigma_{ij}$ is the sign that makes the term $2 p_s k_{\ell(i,j)}$ in
$\sigma_{ij}(\Sigma^2_j -\Sigma^2 _i)$ appear with a positive sign.

\newpage
The full set of functions $f_j$ whose zero's define the locations of the
singularities of the four functions $F_{g}$ represented by the graphs $g$
of Fig. 3
are given in Fig. 4.
The functions $f_j$ for $j= (1, ... , 6)$ correspond to denominators
$f_j + i0$.
The function $f_7$ corresponds to the $\delta$--function constraint $\delta
(\Omega\widetilde{\Omega}-1)$,
and $f_8$ corresponds to the Heaviside function $\theta (r)$.

\begin{figure}
$$
\begin{array}{ll}
(a)&(b)\cr
f_1 = \Omega^2& f_1 = \Omega^2\cr
f_2 = 2p_2 \Omega + r\Omega^2& f_2 = 2p_2\Omega + r \Omega^2\cr
f_3 = (p_2 + r\Omega )^2 - m^2&f_3 = (p_2 + r\Omega )^2 -m^2\cr
f_4  = (p_1 + r\Omega )^2 - m^2&f_4 = 2p_1\Omega + r\Omega^2\cr
f_5 = 2p_1 \Omega + r\Omega^2 &f_5 =  p_1^2 - m^2\cr
f_6 = p^2_3 - m^2 &f_6 = p^2_3 - m_2\cr
f_7 = \Omega{\widetilde\Omega}-1 & f_7 = \Omega\widetilde{\Omega}-1\cr
f_8 = r & f_8 =r\cr
&\cr
(c)&(d)\cr
f_1=\Omega^2&f_1=\Omega^2\cr
f_2=p^2_2-m^2&f_2=p^2_2-m^2\cr
f_3=2p_2\Omega+ r\Omega^2&f_3= 2p_2 \Omega + r\Omega^2\cr
f_4=(p_1+r\Omega)^2-m^2&f_4 =2p_1\Omega+ r\Omega^2\cr
f_5=2p_1\Omega + r\Omega^2&f_5=p^2_1-m^2\cr
f_6 = p^2_3-m^2&f_6=p^2_3-m^2\cr
f_7 = \Omega\widetilde{\Omega} -1&f_7 = \Omega\widetilde{\Omega}-1\cr
f_8 =r&f_8=r\cr
\end{array}
        $$
\caption{
 The functions $f_j$ whose zeros define the singularity surfaces of the
four functions $F(g)$ represented by the four $*$ graphs $g$ of Fig. 3.
Here, and in what follows, the vectors $p_s$, $s\in \{1,2,3\}$, are the vectors
defined beneath Fig. 1.
}
\label{fig4}
\end{figure}

The necessary (Landau--Nakanishi) conditions$^{3,4}$  for a singularity
(in the original real domain of definition) of one of these functions $F_g$ is
that there be a set of real numbers $\alpha_1, ... , \alpha_8$, not all zero,
a real number $ r \geq 0$ ($r \leq \delta$), and a pair of real four--vectors
$\Omega$ and $p$, with
$p_1 =p+q_1, \ \ p_2=p-q_3,$ and $p_3 = p,$ such that
$$
 \alpha_j f_j = 0 \hbox{\hskip .5in all} \ \ \ j\epsilon \{ 1, ... , 8\},
\eqno(4a)
$$
and
$$
\sum^8_{i=j} \alpha_j {\partial f_j\over \partial x_i} =0 \hbox{\hskip .5in
all} \ \ \ \ i\epsilon \{1, 2, 3\},\eqno(4b)
$$
where $x_1 = \Omega, x_2 = r, x_3 = p,$ and
$$
 \alpha_j \geq 0 \hbox{\hskip .5in} j\epsilon \{1, ...,6\}.\eqno(4c)
$$
Also,
$$
f_7 = 0, \ \ \hbox{and} \ \ \ r << m.\eqno(4d)
$$
The contribution from the upper end points of the r integrals are neglected
because these end points are artificially introduced, and hence do not
represent singularities of the full function.
\newpage

The Landau matrix $L_{ij} \equiv \half \partial f_j/\partial x_i$ for the
function represented by the graph of Fig. 3a is shown in Fig. 5.
The Landau (loop) equations (4b) are formed by multiplying each row $j$ of
this matrix by
$\alpha_j$ and requiring the sum of each of its columns to vanish.

\begin{figure}
$$
\begin{array}{lccc}
{f_j}&{d\Omega}&{dr}&{dp}\cr
\hline
f_1=\Omega^2&\Omega&0&0\cr
f_2=2p_2\Omega+r\Omega^2 &p_2+r\Omega &\half\Omega^2&\Omega\cr
f_3= (p_2+r\Omega )^2-m^2&r(p_2+r\Omega )&(p_2+r\Omega )\Omega&p_2+r\Omega\cr
f_4=(p_1 + r\Omega )^2 - m^2&r(p_1+r\Omega )&(p_1+r\Omega
)\Omega&p_1+r\Omega\cr
f_5 = 2p_1 \Omega + r\Omega^2 &p_1+r\Omega  & \half \Omega^2&\Omega\cr
f_6 = p_3^2-m^2 &0&0&p_3\cr
f_7=\Omega\widetilde{\Omega}-1&\widetilde{\Omega}&0&0\cr
f_8=r&0&\half&0\cr
\end{array}
      $$
\caption{
 The Landau matrix $L_{ij}$ corresponding to the graph in Fig. 3a.
The $\sigma_{js}$'s are negative for $j=2$ and $j=5$.}
\label{fig5}
\end{figure}

There are two cases: $r\neq 0$, and $r=0$.
If $r\neq 0$ then the equation (4a) implies $\alpha_8=0$.
If one forms the combination of columns $\Omega d\Omega - rdr$ and compares the
entries to equation (4a),
$\alpha_jf_j=0$, then one finds that the only term in the resulting loop
equations is $\alpha_7\Omega\widetilde{\Omega}=0$,
with $\Omega\widetilde{\Omega}=1$
This entails $\alpha_7 =0$.
If, on the other hand, $r=0$ then the $dr$ column of $L_{ij}$ has an entry in
row 8, and hence it cannot be used in this way.
But for $r=0$ this column does not contribute to $rdr.$
So in either case the conclusion holds: $\alpha_7=0$, and the
$\Omega\widetilde{\Omega}=1
$ row does not contribute.

Similar arguments in the case of graphs with more lines show that one can
always eliminate all of the rows
corresponding to $\Omega_i\widetilde{\Omega}_i-1$.
In the general case it is the combination of columns $\Omega_id\Omega_i -
r_idr_i+r_{i+1}dr_{i+1}$
that is used to show the vanishing of the row corresponding to
$\Omega_i\widetilde{\Omega}_i=1$.
(See Appendix A.)

Consider now the function corresponding to the graph in  Fig. 3d, and the
corresponding set of functions $f_j$ in Fig. 4d.
This graph is a  graph of the separable kind:
cutting the three $*$ segments separates it into three
disjoint parts.

If one considers the $d\Omega$ column with the $\Omega\widetilde{\Omega}=1$
row deleted then one immediately concludes from a look at Fig. 4d, and from the
nonparalled character of $p_2+r\Omega$, and $p_1 + r\Omega$, and the
impossibility of the simultaneous vanishing of $f_1$ and either $f_3$ or $f_4$,
that the only solution of the implied $\Omega$ loop equation [and (4a)] is
the trivial one in
which all three contributions are zero: $\alpha_1=\alpha_3=\alpha_4=0$

In this situation we may invoke a basic lemma$^{7}$:
``For any sets of real numbers $\eta_{ba}$ and $\lambda_{ca} $ the system of
equations
$$
\sigma_b = \sum_a \eta_{ba}\delta_a \hbox{\hskip .25in}\sigma_b  >0 \eqno(5a)
$$
$$
0= \sum_a \lambda_{ca}\delta_a
$$
has a solution $\delta \equiv \{\delta_a\}$ if and only if the system of
equations
$$
\sum_b \alpha_b\eta_{ba} + \sum_c
\beta_c\lambda_{ca} =0 \hbox{\hskip .25in}
\alpha_b \geq 0,\  \sum \alpha_b > 0\eqno(5b)
$$
has no solution $(\alpha , \beta )$.''

Identifying $(\eta_{ba}, \lambda_{ca})$ with the entries in the $d\Omega$ and
$dr$ columns of $L_{ij}$, with $b=j\epsilon \{1, ... . 6\}$ and $c=j\epsilon
\{7, 8\}$, and identifying $\delta_a = \delta\Omega_a$, for $a\epsilon \{ 0, 1,
2, 3\}$, as an imaginary displacement of the four--vector
contour--of--integration variable $\Omega$, we find from this lemma, and the
above--mentioned fact [that the only solution of these equations is the trivial
one with every term equal to zero],
that at every point in the space of integration variables $p$ and $\Omega$
where some set of functions $f_j$ vanishes there is a displacement of the
contour in $\Omega$ space that shifts the contour away from every
$\Omega$--dependent vanishing $f_j$:
by virtue of $(\partial f_j/\partial\Omega )  \delta \Omega >0$ [i.e., (5a)]
every
such function $f_j(\Omega )$ is shifted by this distortion into its upper--half
plane.
\vskip 7pt

\begin{figure}
\caption{ The graph representing a term obtained by pole decomposition.
This graph  separates into three disjoint parts when one cuts the three $*$
segments.}
\epsfxsize = 6.01in
\epsfysize = 3.58in
\epsffile{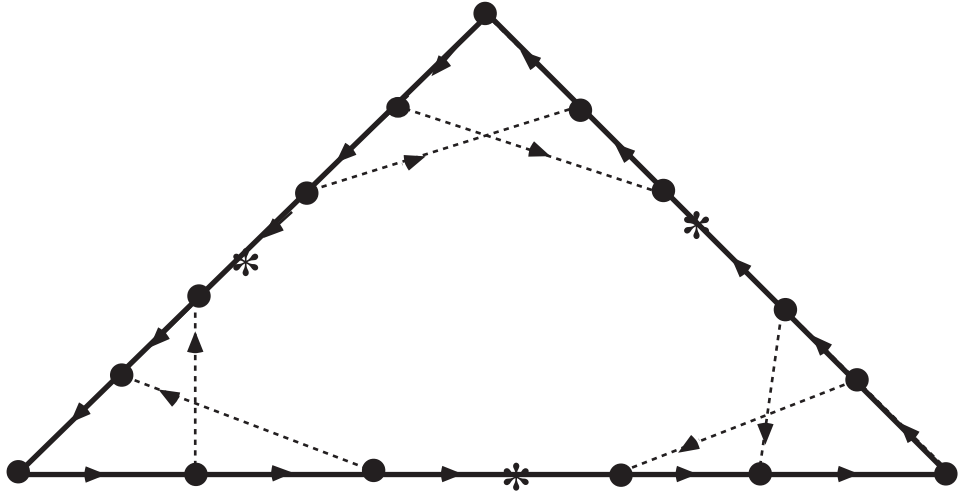}
\label{fig6}
\end{figure}

We wish to generalize this result.
We are particularly interested in the functions represented by separable
graphs,
i.e., by graphs that separate into three disjoint parts when the three
$*$ segments are cut. Another example of such a graph is shown in
 \ref{fig6}

\begin{figure}
\caption{
 The Landau diagram associated with the graph of Fig. 6.
We distinguish `Landau diagrams' from `graphs: the former are geometric, the
latter topological.}
\epsfxsize = 4.52in
\epsfysize = 5.89in
\epsffile{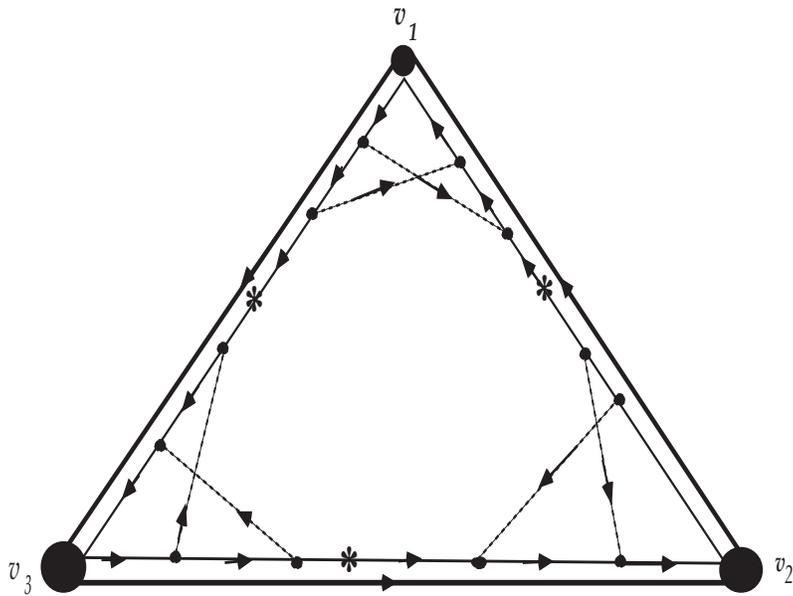}
\label{fig7}
\end{figure}

Consider first the case where all $r_i\neq 0$. In this case the Landau
equations are equivalent to the Landau equations that arise from using the
k-space variables, instead of the $(r,\Omega)$ variables.
Then the Landau equations associated with the function represented by the
graph shown in Fig. 6 can be expressed in a simple geometric form:
these equations  are equivalent to the existence of a
 ``Landau diagram'' (a diagram in
four--dimensional space) that has the form shown in \ref{fig7}.

This Landau diagram is a diagram in four-dimensional space (thought of as
spacetime),
and each segment of the diagram represents a four vector.
The rules are these:
\begin{enumerate}
\item Each directed photon line segment  $i$ represents the vector
$$
V_i = \alpha_i k_i,\eqno(6a)
$$
where $k_i$ is the momentum flowing along segment $i$ of the graph in
the direction of the arrow, and $\alpha_i \geq 0$.
\item Each directed charged-particle segment $j$ corresponding to a
pole-residue factor $f_j$ represents the vector
$$
V_j = \beta_{js} \Sigma_j, \eqno(6b)
$$
where $\sum_j$ is the momentum flowing along segment $j$ of the graph in the
direction of the arrow on it, and
$$
\sigma_{js}\beta_{js} =\alpha_j \geq 0,\eqno(6c)
$$
where the sign $\sigma_{js}$ is defined below (3).
\item Each directed charged-particle line segment $s$ corresponding to a pole
denominator $(\sum^2_s - m^2 + i0)$ is represented by
a star (asterisk) line segment $s$, and it represents the vector.

$$
V'_s = \alpha^\prime_s \Sigma_s \ \ \ \ \ ,\eqno(6d)
$$
where $\sum_s$ is the momentum flowing along $*$ line segment $s$ of the graph
in
the direction shown, and
$$
\alpha'_s = \alpha_s -\sum_{j\epsilon J(s)} \beta_{js}.\eqno(6e)
$$
Here $\alpha_s$ is the Landau parameter $\alpha$ corresponding to the
function $f_s = \sum^2_s - m^2 +i0$, and for each side $s$ the
set $J(s)$ is the set of indices $j$ that label the pole-residue denominators
that are associated with side s of the triangle graph.
\item Three line segments appear in the Landau diagram that are not images of
segments that appear in the graph.
They are the three {\it direct} line segments that directly connect pairs of
vertices from the set $\{ v_1, v_2, v_3\}$.
The vector $V_s$ associated with the direct segment $s$ is
$$
V_s = \alpha_s\Sigma_s + \sum_{j\epsilon J(s)} \beta_{js}
(\Sigma_j - \Sigma_s).\eqno(6f)
$$
It is equal to the sum of the vectors corresponding to the sequence of $*$ and
non $*$ charged-particle line segments that connect the pair of vertices $ v_i
$
between which the direct line segment $s$ runs.
\end{enumerate}

The $p$ loop equation is represented by the closed
loop formed by the three direct line segments $V_s$ specified in (6f)
The photon loop equation associated with the photon line carrying momentum
$k_i$ is formed by adding to $\alpha_i k_i$
the sum of the vectors corresponding to the charged-particle segments needed
to complete a closed loop in the diagram (See
Appendix B).
Thus the existence of a (nontrivial) solution of the Landau equations is
equivalent to the existence of a (nonpoint) Landau diagram
having the specified topological structure, with its line segments
equal to the vectors specified in (6).

Although Figs. 6 and 7 represent a separable case the rules described
above general: they cover all cases in which all $r_i$ are nonzero.

For each s we can use in the Landau diagram either $V'_s$ or $V_s$. We shall
henceforth use always
$V_s$, the segment that directly connects a pair of vertices $v_i$, rather than
$V'_s$, and we shall place a star (asterisk) on each of these three direct
line segments. These three direct line segments are geometrically more useful
than the $V'_s\/$'s because they display immediately the p loop equations, and
also the relative
locations of the three external vertices $v_i$, and because each one has only a
single contribution, $\alpha_s\Sigma_s$, of well-defined sign and direction,
in the limit $k_i \Rightarrow 0$, provided condition (8)(see below) holds.

We specify the way that photon loops pass through Landau diagrams:
a photon loop shall pass through the star
line $s$ of a {\it Landau diagram} (i.e., along the direct line segment $s$)
if and only if the corresponding loop in the {\it graph\/}
passes through the star line $s$ of the {\it graph}.

The positivity of the photon-line $\alpha_i$'s entails that each directed
vector
$\alpha_ik_i$ of Fig. 7 points in the positive (energy/time)
direction (i.e., to the left) if the energy $k^0_i$ is
positive, and in the negative direction (i.e., to the right) if the energy
$k^0_i$
is negative.
This fact entails that positive energy is carried by each nonzero (
length)
photon line
segment of Fig. 7 {\it out of} the vertex that stands on its right--hand end
and {\it into} the vertex that stands on its left--hand end.
This result is true independently of the direction in which the arrow points,
or of the sign of the energy component $k_i^0$.

In the general separable case some of the non $*$
segments may have $\alpha_j = 0$, and hence contract to points.
Consequently several photons may emerge from, or enter into, a single vertex
of the Landau diagram.

This geometric representation of the ``Landau'' equations holds only if all
$r_i \neq 0$. If one or more $r_i=0$ then the diagram breaks into parts, as
will be seen. We wish to show, by using these geometric conditions and the
result (5), that the $\Omega_i$ contours can be distorted in such a way as to
avoid simultaneosly all the singularities except those associated with the
three $*$ line poles, one for each of the three sides $s$ of $G$, and
those associated with the various end points $r_i=0$ and $r_i=1$.
We shall treat the various cases separately.

\newpage

\noindent {\bf 3. Separable Case; All $r_i\neq 0$.}

To prove this result for the separable  case, and when all $r_i\neq 0$, let us
consider any one of the three disjoint partial diagrams of non $*$ segments.
Let $V$ be the set of vertices of this partial diagram that lie on an
end of at least one photon line that is not contracted to a point.
Let $V_R$ be any element of $V$ such that every nonzero-length photon line
incident upon $V_R$ has its other end lying to the left of $V_R$. Let $V_L$
be any element of $V$ such that every nonzero-length photon line that is
incident upon $V_L$ has its other end lying to the right of $V_L$.
Then the total momentum $K$ carried into either $V_R$ or $V_L$ by
all photons incident upon it satisfies $K\neq 0$ and $K^2 \geq 0$:
these properties  follow from the fact that each photon line of nonzero
length incident upon $V_R$ must carry a light-cone-directed momentum-energy
with positive energy out of $V_R$, and each photon line
of nonzero length incident upon $V_L$ must carry a light-cone-directed
momentum-energy with positive energy into $V_L$.
However, one cannot satisfy $2pK + K^2=0$ with
$p\simeq p_1, p_2$ or $p_3$,
and with a small $K\neq 0$ satisfying $K^2\geq 0$.
Consequently the charged--particle line segments of the partial Landau diagram
lying on the outer
extremities of the two charged particle lines must contract to points, by
virtue of (4a): the associated Landau parameter $\alpha_i$ must vanish.
Recursive use of this fact entails that {\it all}
of the lines in this partial diagram must contract to a single point.

The existence of zero--length photon lines whose ends do not lie in $V$
does not disturb this argument, provided self--energy parts are excluded.

This result, that each non $*$ line contracts to a point, means that every
entry
in every $\Omega_i$ loop equation vanishes.
Under this condition the lemma expressed by Eq. (5) shows that
every $\Omega_i$ contour can be
distorted away from every $\Omega_i$--dependent singularity.

We next show that this result continues to hold when some or all of the
$r_i$ vanish.

\vskip 9pt
\newpage
\noindent {\bf 4. Separable Case; Some $r_i=0$}

\begin{figure}
\caption{
 Part of the diagram of Fig. 7.}
\epsfxsize = 4.52in
\epsfysize = 5.89in
\epsffile{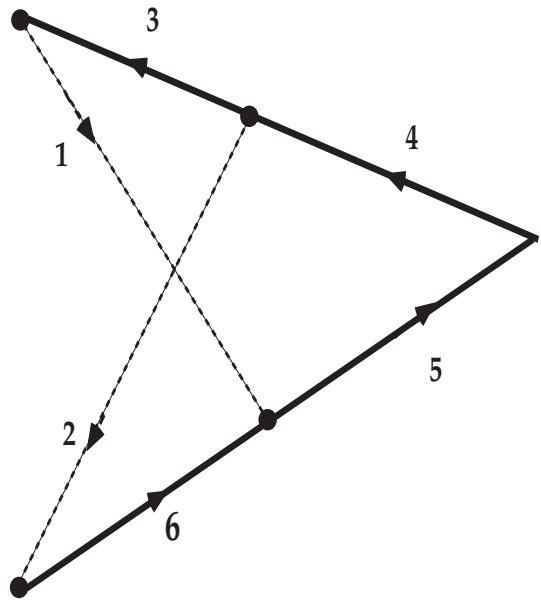}
\label{fig8}
\end{figure}

Let us first consider the simple example shown in \ref{fig8}.

The Landau matrix for the diagram of Fig. 8 is shown in Fig. 9.

\begin{figure}
$$
\begin{array}{lcc}
 f_j&{d\Omega_1}&{d\Omega_2}\cr
\hline
&&\cr
\Omega^2_1&\Omega_1&0\cr
&&\cr
\Omega^2_2&0&\Omega_2\cr
&&\cr
2p_1\Omega_1+r_1\Omega^2_1&p_1+r_1\Omega_1&0\cr
&&\cr
2p_1(\Omega_1+r_2\Omega_2)+r_1(\Omega_1+r_2\Omega_2)^2&p_1+r_1\Omega_1+r_1
r_2\Omega_2&r_2(p_1+r_1\Omega_1+r_1r_2\Omega_2)\cr
&&\cr
2p_2(\Omega_1+r_2\Omega_2)+r_1(\Omega_1+r_2\Omega_2)^2 \ \ & p_2+r_1
\Omega_1+r_1r_2\Omega_2\ \ \ &r_2(p_1+r_1\Omega_1+r_1r_2\Omega_2)\cr
&&\cr
2p_2\Omega_2+r_1r_2\Omega_2^2&0&p_2+r_1r_2\Omega_2\cr
\end{array}
$$
\caption{
 The Landau matrix corresponding to the diagram of Fig. 8.
The rows corresponding to the conditions
$\Omega_j\widetilde{\Omega}_j=1$ have been removed, by using the
argument given in Appendix A.}
\label{fig9}
\end{figure}

If $r_1\neq 0\neq r_2$ then one can multiply the $\Omega^2_1$ row by $r_1$,
multiply the $\Omega^2_2$ row by $r_1r^2_2$,
multiply the last row by $r_2$, and divide the $d\Omega_2$
column by $r_2$.
This brings the matrix into an equivalent one in which $r_1$ and $r_2$ occur
only in the combinations $k_1=r_1\Omega_1$ and $k_2=r_1r_2\Omega_2$: this is
the equivalent $k$ form that was previously  used for the case $r_1\neq
0\neq r_2$.

If $r_1=0$ and $r_2\neq 0$ then one can perform the same transformations
involving $r_2$, and bring the equations to the same form as before, except
that the vector associated with the photon line segment 1 is now
$\alpha_1\Omega_1$
instead of $\alpha_1k_1$, and the vector associated with the photon line
segment 2 is now $\alpha_2 r_2 \Omega_2$ instead of
$\alpha_2 k_2$.
The vectors $r_1\Omega_1$ and $r_1r_2\Omega_2$ that occur summed with $p_1$
or $p_2$ become zero.
Thus the situation is geometrically essentially the same as in the case
$r_1\neq 0\neq r_2$,
though slightly simpler:
the small additions $k_1$ and $k_2$ to the vectors $p_1$ and $p_2$ now
drop out.
The important point is that the critical denominators $2pK+K^2$ of the
earlier argument now take the form $2p\Omega$, with $\Omega^2 \geq 0$ and
$\Omega \neq 0$.
Such a product cannot vanish.
Thus the earlier $r_i \neq 0$ argument goes through virtually unchanged.

If $r_1\neq 0$ and $r_2=0$ then the $\Omega_1$ and $\Omega_2$ loop equations
can be considered separately.
The earlier $r_i \neq 0$ argument of section 3 can be applied to
the  first part alone, and it shows that each line segment
on the $\Omega_1$ loop must contract to a point.
Next the $\Omega_2$ equation can be considered alone, with
each segment along which the $\Omega_1$ loop flows contracted to a point.
Then the earlier $r_1 = 0$ arguments can be applied now to this
$\Omega_2$ part of the
diagram (with $r_2$ in place of $r_1$): it shows that each of the segments
along which $\Omega_2$ flows also must  contract to a point: the corresponding
$\alpha_j$ must be zero.

The case $r_1 = r_2 = 0$
is not much different from the case just treated: $r_1$ enters Fig. 9
only in an unimportant way.

The generalization of this argument from the case of Fig. 9
to the  general separable case is straightforward.
Let $r_g$ be the first vanishing element of the ordered
set $r_1, r_2, ..., r_n$.
Then the set of  $\Omega$ columns of the  Landau matrix
separates into one part involving only the $\Omega_i$ columns for $i <
g$, and a second part involving only the
$\Omega_i$ columns for $i\geq g$.
For the first part of this matrix the argument given above for the case with
all $r_i\neq 0$
holds, and it entails that every line segment in this part must contract to a
point. With all of the rows corresponding to these contracted segments omitted
one may apply the
$r_1=0$ argument (with $r_g$ in place of $r_1$) to the part
$i\geq g$, and proceed iteratively.
This arguments leads to conclusion that the only solution to all of the
$\Omega_i$ loop equations is
the trivial one where every entry in every $\Omega$ column  is zero.
Hence the lemma expressed by Eq. (5)  ensures that each $\Omega_i$ contour
can be distorted away from all of its singularities, in the general separable
case.

As one moves from the domain where all $r_i>0$ to the various boundary points
where some $r_i=0$
\ \ two kinds of changes can occur.
Certain conditions that particular vectors $\Omega_j$ be in the upper-half
plane
with respect to a variable like
$(p_1+r_1\Omega_1+r_1r_2\Omega_2) \cdot \Omega_j$
becomes slightly simplified when an $r_i$ becomes zero.
Since the different conditions of this kind  correspond to vectors $p_1, p_2$,
and $p_3$ that are well separated, the passage to a point $r_i=0$ causes no
discontinuous change in the set of vectors that satisfy such conditions.
The second kind of change is that some contributions to particular
$d\Omega_j$'s may suddenly drop out if some $r_i$ vanishes.
(See Fig. 9 with $r_2=0$).
These changes at the boundary points of the region $r_i  \geq 0$ do not entail
any discontinuity in the distortion of the $\Omega $ contours on the boundary.
The possibility of using a distortion in $\Omega $ space that is everywhere
continuous in $(r,\Omega)$ follows from the continuousness of the gradients of
the functions $f_i(r,\Omega)$, and the fact that at every point in the domain
of
integration the set of gradients of the set of vanishing $f_i$ form a {\it
convex} set:  the Landau equations cannot be satisfied.

\newpage

\noindent{\bf 5. Nonseparable Case; All $r_i\neq 0.$}

We consider next the functions represented by graphs such that the cutting of
the three $*$ segments does not separate the graph into three disjoint
parts.
The same result about distortions of $\Omega_i$ contours can be obtained also
for these functions.

To obtain this result we consider first, as before, the case in which all
$r_i\neq 0$.
Then we may use the $k$ form of the Landau equations given in (6).

The argument proceeds as before, by making use of the vertices $V_R$ and $V_L$.
No such vertex can join together two pole-residue segments $j$ of
nonzero length: it is impossible to satisfy both $2pK_1+K_1^2=0$ and
$2pK_2+K^2_2=0$ if $K_1-K_2=K$ satisfies $K^2 \geq 0$ and $K\neq 0$, and
$K_1$ and $K_2$ are small compared to the timelike $p$.
Likewise, neither $V_R$ nor $V_L$ can join a $*$ segment to a pole-residue
segment $j$ with $\alpha_j \neq 0$:
one cannot satisfy $2pK+K^2 = (2p+K)K=0$
if $K^2\geq 0$ and $K\neq 0$, and $K$ is much smaller than the timelike
$p$.
Consequently each of the vertices $V_R$ and $V_L$ must be confined to the set
of
external vertices $v_i$:
$$
\{ V_R, V_L\} \subset \{ v_1, v_2, v_3\}.\eqno(7)
$$
\vskip 9pt

\begin{figure}
\caption{
 The Landau diagram corresponding to the $*$ graph (c) of Fig. 3.
This diagram represents the equations obtained from Fig. 6c, with $f_1$
multiplied by
$r^2$, $f_3$ and $f_4$ multiplied by $r$, and $r\Omega$ replaced by $k$.
These changes recover the $k$ form of the equations.
The backward orientation of the vector $\alpha_5 p_1$ arises from the
negative sign of $\sigma_{51}$. However,
this vector is oriented against the direction of the photon loop. Consequently
all contributions to this photon-loop equation proportional to any
 $p_s$ have the form $\alpha_j p_s$: the two reversals of the line segment
$j=5$ compensate for each
 other.}
\epsfxsize = 4.52in
\epsfysize = 2.89in
\epsffile{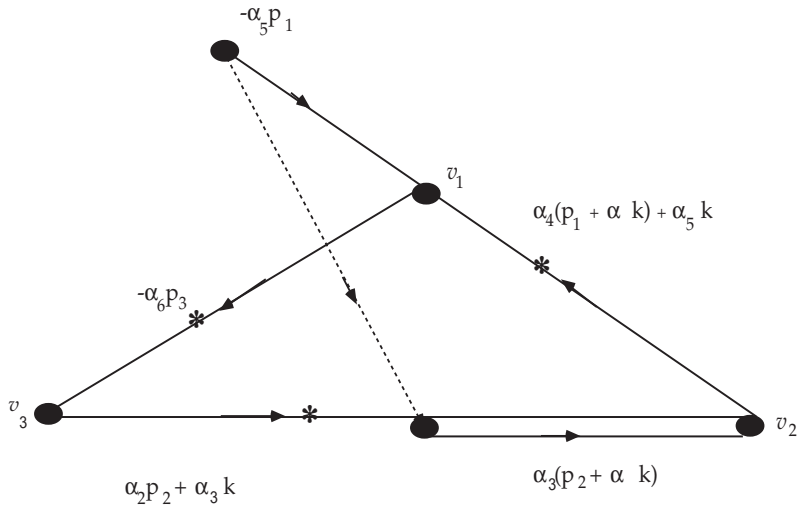}
\label{fig10}
\end{figure}

In the nonseparable case some of the signs $\sigma_{js}$ will be negative.
Consequently some of the vectors corresponding to pole-residue factors $f_j$
will point in the `reversed' direction, because their $\beta_{js}\/$'s, defined
in (6c), are negative.
There are also some (sometimes-compensating) reversals of the ways that certain
photon loops run. These latter reversals arise because we have used, in the
Landau diagrams, the three line segments
that directly connect the pairs in $\{ v_1, v_2, v_3 \}$,
rather than the images of the three star lines of the original $*$ graph.
For example, the $*$ graph of Fig. 3c gives a Landau diagram of the form
shown in \ref{fig10}.

\begin{figure}
\caption{Figur A graph representing a term in the pole-decomposition
expansion.}
\epsfxsize = 4.87in
\epsfysize = 2.39in
\epsffile{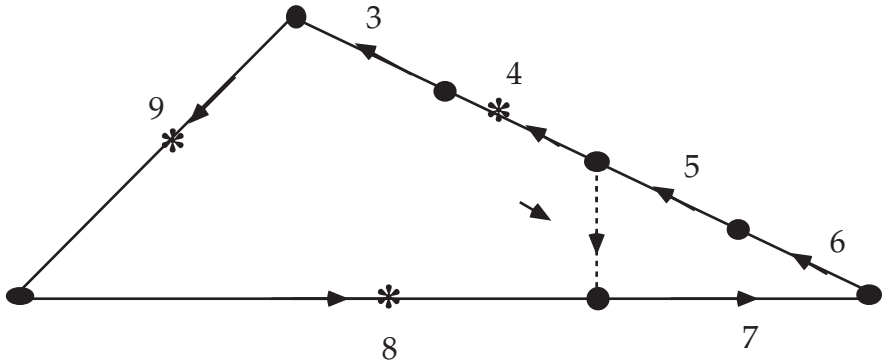}
\label{fig11}
\end{figure}

A second example is the function represented by the graph shown in
\ref{fig11}.

The functions $f_j$ and the Landau matrix  corresponding to the function
represented by the graph  in Fig. 11 are shown in Fig. 12, for $\abs{k_1}>
\abs{k_2}>0$

\begin{figure}
   $$
\begin{array}{lccc}
{f_j}&{dk_1}&{dk_2}&{dp}
\cr
\hline
f_1=k^2_1&k_1&0&0\cr
f_2=k^2_2&0&k_2&0\cr
f_3=2p_1k_1+k^2_1&p_1+k_1&0&k_1\cr
f_4=(p_1+k_1)^2-m^2&p_1+k_1&0&p_1+k_1\cr
f_5=2p_1k_2 +2k_1k_2+k^2_2&k_2&p_1+k_1 + k_2&k_2\cr
f_6=2p_1(k_1-k_2)+k^2_1-k^2_2&p_1+k_1&-(p_1+k_2)&k_1-k_2\cr
f_7=2p_2k_2 + k_2^2 &0&p_2+k_2&k_2\cr
f_8=p^2_2-m^2&0&0&p_2\cr
f_9=p^2_3-m^2 &0&0&p_3\cr
\end{array}
$$
\caption{ The Landau matrix for the function represented by the
graph in Fig.
11, for $\abs{k_1}>\abs{k_2}>0$.
The sign of $\sigma_{j1}$ is minus for $j=3$ and 6,
and otherwise plus.}
\label{fig12}
\end{figure}

\begin{figure}
\caption{
 The `Landau diagram' that represents the Landau equations
associated with the Landau matrix shown in Fig. 12. This diagram is not a true
Landau diagram, because, for example, the vector $\alpha_i k_i$
cannot be a light-cone vector. Moreover, condition (7) is not
satisfied. Were it not for the non-negativity condition on $\alpha_3$
one could satisfy the Landau equations with $\alpha_3 =- \alpha_4$, and
$ \alpha_1 = \alpha_2 =\alpha_5 = \alpha_6 = \alpha_7 =0$.}
\epsfxsize = 7.06in
\epsfysize = 4.37in
\epsffile{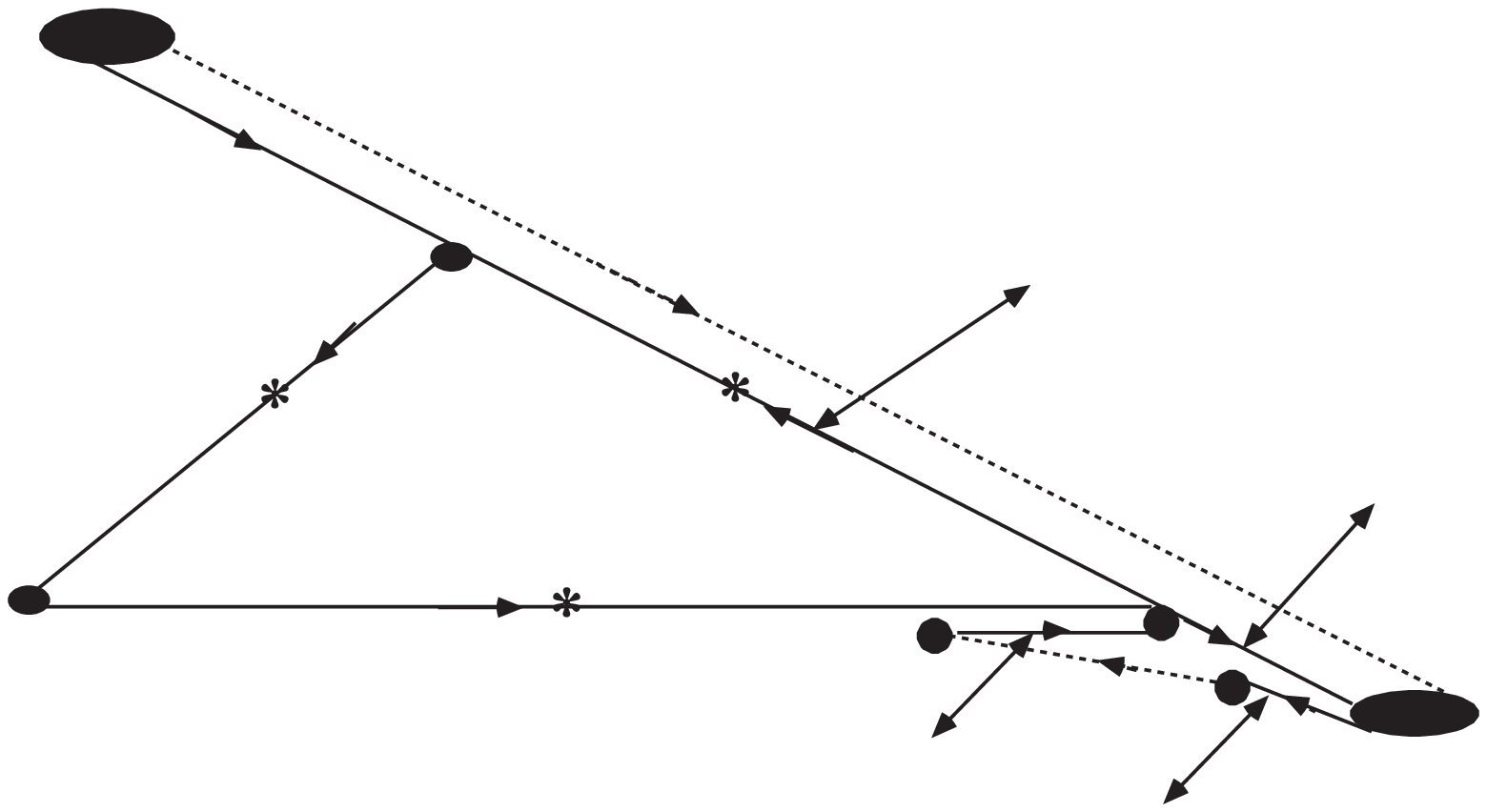}
\label{fig13}
\end{figure}

The Landau diagram corresponding to the Landau matrix in Fig. 12 is shown in
\ref{fig13}

The argument leading to (7) entails more than (7).
It shows, in the present case where all $k_i\neq 0$,
that each vertex of the diagram that does not lie in $\{ v_1, v_2, v_3\}$ and
that has
at least one nonzero-length photon line segment incident upon it must have
 at least  two nonzero-length photon lines incident upon it: each such vertex
must lie on the right-hand end of at least one such photon line segment,
and on the left-hand end of some other such photon
line segment.
Consequently, every nonzero-length photon line must lie on a `zig-zag'
path of photon lines that begins at a vertex in the set
$\{v_1, v_2, v_3\}$,
moves always to the left, and ends on another vertex in $\{v_1, v_2, v_3\}$:
 only in this way can the conditions $K^2 \geq 0$ and $K\neq
0$ used in the derivation of (7) be overcome, if all $k_i$ are different from
zero.

\begin{figure}
\caption{
  A triangle graph with photon vertices labelled by numbers, and
charged--particle line segments labelled by letters.
The segments $h, c,$ and $n$ are  $*$ segments associated with the
pole-decomposition formula $(0')$.
The photon lines have been suppressed.}
\epsfxsize = 6.52in
\epsfysize = 3.50in
\epsffile{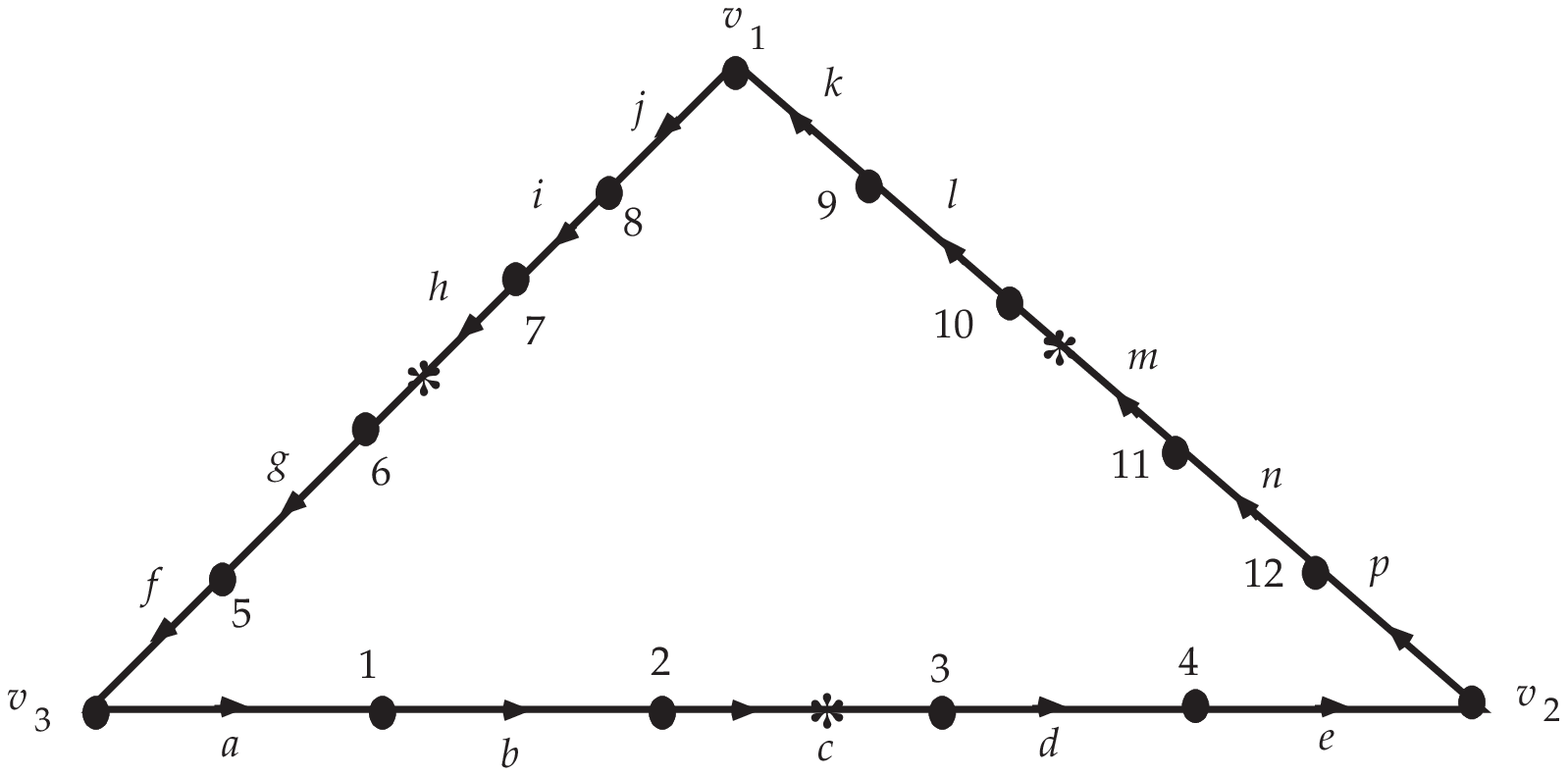}
\label{fig14}
\end{figure}

Consider, then, an example with vertices labelled as in \ref{fig14}.

Suppose $V_L=v_3$ and $V_R=v_1$ are the unique $V_L$ and $V_R$.
Then some sequence of photon lines of nonzero length
must join together to  give a zig--zag path from $v_1$ to $v_3$.
Three examples are shown in \ref{fig15.}

\begin{figure}
\caption{
 Three diagrams with zig--zag paths of photons connecting $v_1$ to
$v_3$.}
\epsfxsize = 6in
\epsfysize = 6in
\epsffile{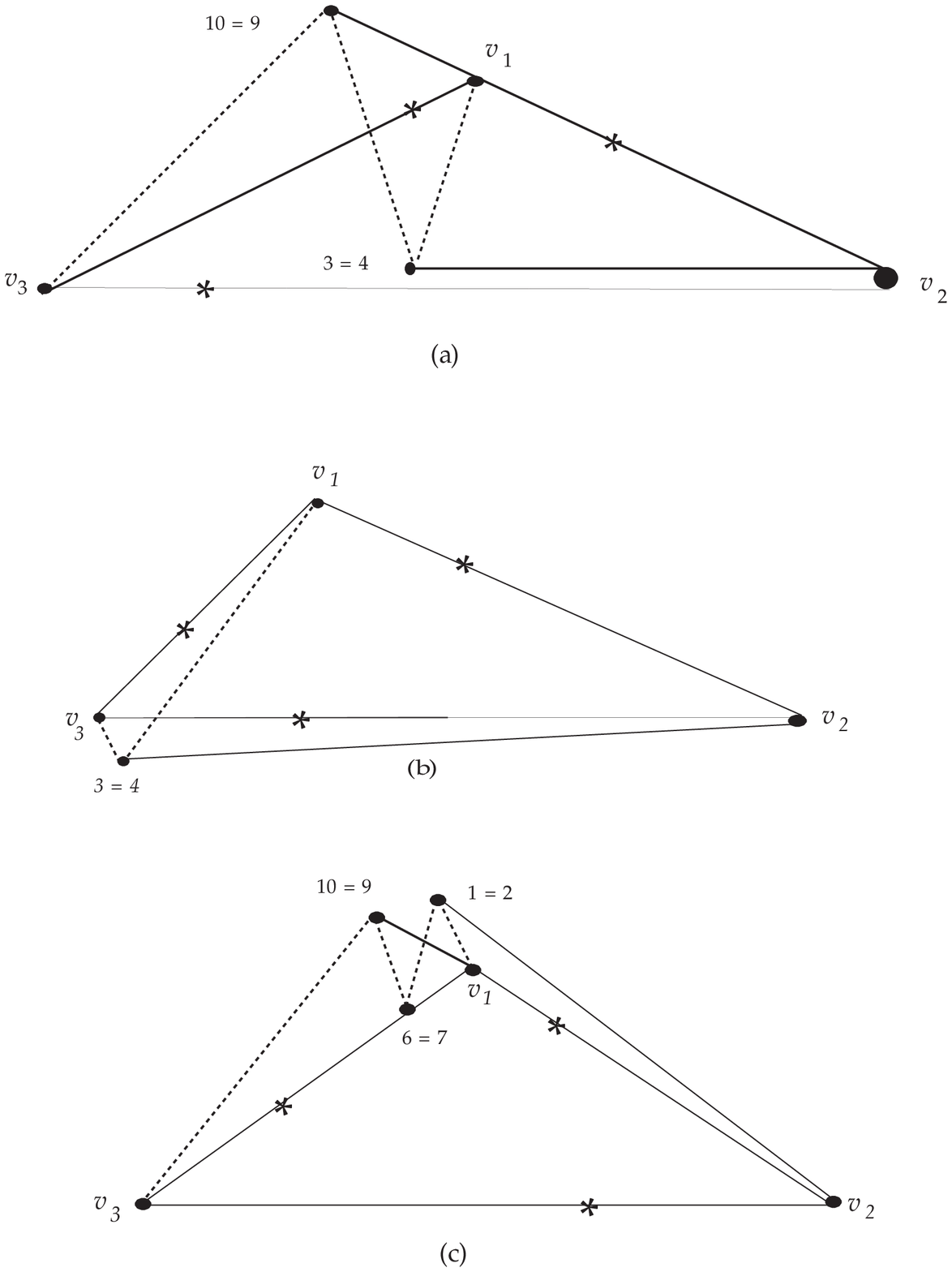}
\label{fig15}
\end{figure}

To analyse such diagrams we assume temporarily that for all pertinent
solutions of the Landau equations
$$
|\alpha_j | \leq |\alpha_s |B\   \hbox{for}\ j\epsilon J(s),\eqno(8)
$$
where $B$ is some fixed finite number. That is, we exclude temporarily the case
where some $\alpha_j$ becomes unbounded, with the $\alpha_s$ bounded.
Then as one lets the $\delta '$ in (2) tend to zero the vector
$V_s$ defined in (6f) and, for $j \in J(s)$, the vectors
$V_j$ defined in (6b) all become increasingly parallel
to $p_s$.

Consider then a sequence of bounds $\delta '_t$, $ t=1,2, ... $, that tend to
zero, and a corresponding sequence of solutions $S_t$ to the Landau
equations in which:\\
1) $k_i \neq 0, \ i=1, ..., n$;\\
2) $|k_i| \leq \delta'_t/n, \ i=1, ... , n$;\\
3) some $\alpha_ik_i\neq 0$; and\\
4) condition (8) holds.\\
If $q^t=(q^t_1, q^t_2, q^t_3)$ is the vector $q=(q_1, q_2, q_3)$
specified by $S_t$, then any accumulation
point $\overline{q}$ of the set $\{ q^t\}$ must be specified by a limiting
diagram in which every
charged-particle segment is parallel to one of the vectors $p_s, s\in
\{1,2,3\}$, and in which some zig-zag path of light-cone vectors runs leftward
from a
vertex $V_R$ of $\{ v_1, v_2, v_3\}$ to a vertex $V_L$ of $\{ v_1, v_2,
v_3\}$, but carries zero momentum-energy. The limit point $\overline{q}$
must therefore lie on the Landau triangle diagram singularity surface
$\varphi (q) =0$. However, the presence of the zig-zag photon line connecting
two of the three vertices $v_i$ imposes an extra condition, which define a
codimension-one submanifold of $\varphi (q) =0$.
These submanifolds are finite in number (for any fixed graph $g$), and hence
are nondense in the interior of $\varphi =0$.
If a point $q \in \{ \varphi =0\}$ lies at a nonzero distance from
each of these submanifolds then
no solution of the kind specified above can occur, and hence
for some sufficiently small
neighborhood $N$ of $q$, and for some sufficiently small $\delta '$, any
solution to the Landau equations for $q \in N$ satisfying $0< |k_i|
\leq \delta '/n$ for all $i$, and conditions 1) and 4), can have only
zero-length photon lines: i.e., for all photon lines $i$
$$
\alpha_i k_i = 0\ \ .   \eqno(9)
$$

We are interested here in the singularity structure at a general
point on $\varphi = 0$, rather than at special points where other
singularity surfaces are relevant. Hence we may restrict our attention to a
neighborhood $N$ in $\varphi =0$ where (9) holds.

Condition (9) says that every photon line segment $i$ must have
zero length. This condition entails the stronger result that every segment on
every photon loop $i$ in the Landau diagram must contract to a point.

To obtain  this stronger result consider {\it in order} the loop equations
corresponding to the
sequence of variables $k_1, ... , k_n$, as defined in the formula $(0)$.

Consider first, then, the closed loop 1 in the Landau diagram.
For each charged-particle segment on this loop the $k_{\ell}$ with
smallest $\ell$ that flows along this loop 1 is $k_1$ itself.
Consequently the orientations of all of the segments along this loop
are unambiguously
determined: for each $s \in \{ 1,2,3\}$ every contribution to the loop
1 that arises from a charged-particle segment on side $s$ {\it adds} to the
loop equation a vector that is very close to a non-negative
multiple of $p_s$, just as in Figs. 10 and 13.
Use can be made here of the facts$^{8-11}$ that the triple of four-vectors
$(v_1, v_2, v_3)$
specified by the three external vertices $v_i$ constitute a normal to the
Landau surface (in $q= (q_1, q_2, q_3)$ space)  associated with the diagram,
 and that  this surface can be
tangent to the triangle diagram Landau surface $\varphi (q) = 0$ at a point $q$
only if the directions of the three vectors $V_s$ are the same as they are for
the simple Landau diagram that corresponds to figure 4. Because we are staying
away from exceptional points of lower dimension the three vectors
$V_s$ must be parallel to the three vectors $p_s$. Alternatively, one can use
the condition (8), and take $\delta'$ sufficiently small, in order to deduce
that $V_s$ is approximately equal to $\alpha_s p_s$.

Each photon loop  passes along at most {\it two} sides  s of the
triangle. Hence, on any single photon loop in the
Landau diagram, each charged-particle segment points approximately in the
direction of one or the other of
at most {\it two} of the three vectors $p_s$.
(See Figs. 10 and 13.) Hence the
contraction to a point, demanded by (9), of the remaining segment of the
loop, namely $\alpha_1k_1$, forces every segment on loop 1 to contract to
a point.

Consider next the loop 2.
All segments along which $k_1$ runs have now been contracted out.
Thus the $k_{\ell}$ with the smallest value of $\ell$ that flows along the
surviving part of loop 2 is $k_2$ itself.
Hence each segment on this loop also must contract to a point, by the same
argument that was just used for loop 1. Proceeding step by step one
finds that every segment on every photon loop
must contract to a point.

In this nonseparable case with all $r_i \neq 0$ at
least one photon line must pass along a star line. Hence at least one of the
three star lines of the Landau diagram must also contract to a point. But then
the other two sides of the triangle $(v_1, v_2, v_3)$ must also contract to
points, since, in accordance with the
conditions imposed below Eq. (1), the three sides of the triangle connecting
the three vertices $v_i$ are nonparallel. But then every segment of the
Landau diagram is forced to a
point, and thus there is no solution of the Landau equations, in this
nonseparable case with all $r_i \neq 0$.

This conclusion was derived under the assumption (8). However, that
assumption is not necessary.
Suppose we normalized the solutions by the requiring that max $|v_i -v_j
|=1$, and drop (8).
Then the direction of $V_s$ is not constrained, but its Euclidean  length is.

Consider, under these conditions, the sequence of loops $i$.
A first part of loop 1 consists of either the zero, one, or two vectors
$V_s$ that are included on the loop. Their directions are indeterminate,
but their magnitudes are at most unity.
In fact the magnitude of the sum of these segments is at
most unity.

A second part of this closed loop is the segment corresponding to the
photon 1 itself.
The length of this segment is limited by the fact  that any nonzero-length
photon line segment must lie on a zig-zag path that runs between two of the
vertices $v_i$, and is composed of leftward
pointing light-cone vectors.
Since the Euclidean distance between the endpoints of this zig-zag path is
bounded by unity, the individual segments along this path are likewise bounded.
Thus these first two parts of loop 1 are bounded.

The third and final part of loop 1 is the sum of the contribution of the
segments $j$ associated with the pole-residue denominators $f_j$.
All of these contributions to the loop are essentially of the form
$\alpha_j p_s$, with all the $\alpha_j$'s positive, and $s$
ranging over either {\it one or two\/} of its three possible values.
(See Figs. 10 and 13).
We can impose the condition that at the points $q \in \{ \varphi =
0\}$ under consideration the three vectors $p_s$ are {\it far\/} from parallel.
In this case the bound on the first two parts of the closed loop 1 imposes a
comparable bound on the third part, and, in particular, a bound on the
sum of the $\alpha_j$ corresponding to those segments $j$ that lie on
loop 1.

We then turn to loop 2.
Bounds are established as before for all parts of loop 2 that are not
pole-residue segments $j$, and also for all pole-residue segments $j$ that
lie on loop 1.
Since the contributions from the pole-residue segments $j$ that lie on
loop 2 but not loop 1 have the form $\alpha_j p_s$, with
$\alpha_j\geq 0$, and with $s$
ranging over at most
two of the three possible values, we can now establish
upper bounds on the sum of these new  $\alpha_j$'s.
Proceeding in this way we establish bounds on all of the $\alpha_j\/$'s
associated with all the pole-residue denominators $f_j$.
Then for a sufficiently small $\delta'$ we can ensure that, for
each value of $s$, the contribution to $V_s$, specified by (6f),
that arises from the photon momenta $k_i$ is small compared to this
vector $V_s$ itself.
This is the result that in the earlier argument was obtained from (8),
which we therefore no longer need.

\newpage
\noindent {\bf 6. Nonseparable Case; Some $r_i=0$.}
\vskip 9pt

The results for the $k_i \neq 0$ case carry over to the general situation,
provided the $(r, \Omega)$ variables are retained.

The argument for the case where some  $r_i=0$ proceeds much as in the
case of separable diagrams. Let $r_g$ be the first vanishing member of the
ordered sequence $r_1, r_2,..., r_n$. Then the Landau matrix separates into
two parts. The first consists of the $d\Omega_i$ columns for $i< g$, plus
the $dp$ column; the second consists of the $d\Omega_i$ columns for
$i\geq g$. By multiplying and dividing various rows and columns of the
Landau matrix by appropriate nonzero factors $r_i$ ($i<g$) one can
convert the $i<g$ part to the $k$ form, with all $k_j$ for $j\geq g$
set to zero. The $r_i\neq 0$ argument can then be applied to these $i< g$
Landau equations: they imply the vanishing of the $\alpha_j\/$'s corresponding
to all segments $j$ of the Landau diagram along which run the photon
loops $i$ with $i< g$.

The remaining columns, which give the $i\geq g$ part of the Landau equations,
can be separated into {\it sectors}, where each sector begins with a column
$d\Omega_i$ such that $r_i=0$, and is followed  by the set of columns
$d\Omega_{i+1},..., d\Omega_{i+h}$ such that $r_{i+1},..., r_{i+h}$
are all nonzero. These latter $r\/$'s can be changed to unity without altering
the content of the Landau equations. We shall do this, purely for notational
convenience.

The rows corresponding to the three pole denominators
do not contribute to the $i\geq g$ equations because

$$
\half {\partial\over \partial \Omega_j}[ (p + r_1 ... r_j\Omega_j + ... )^2
-m^2] =0\
\hbox{for}\  j \geq g,$$
due to $r_g=0$.

One proceeds step-by-step, starting with the $i< g$ part, then
considering the various individual sectors, in order of increasing values of
$i$.
The Landau equations for each one of the individual sectors can be expressed by
a Landau diagram constructed in accordance with the rules (6), with, however,
the following changes:
(1), the three vectors $V_s$ corresponding to the three direct line segments
$s$ are set to zero; (2), all the segments of the Landau diagrams that
occur at earlier stages of the step-by-step process are contracted to points;
and (3), the photon propagator contribution  $\alpha_i k_i$  to each
$d\Omega_i$ column  that belongs to the sector in question is replaced by
$\alpha_i \Omega_i$.

The Landau diagram corresponding to a sector S
has a `spider' form: it consists of a single central vertex $v$, which
represents the three coincident vertices $v_i$,
plus a web of segments sprouting out from $v$.
All segments of the Landau diagrams corresponding to the previously considered
sectors are contracted
to the single point $v$, together with all of the segments that  constitute the
part $i<g$.  All charged-particle segments of
the Landau diagram along which run {\it none\/} of the photon loops
that constitute S are also contracted to points.

The Landau diagram that corresponds to any individual sector S can
be shown to contract to a point by using the arguments developed earlier:
the argument involving $V_R$ and $V_L$ shows that no photon line of nonzero
length can occur in the spider diagram, and then the step-by-step
consideration of the photon loops $i$, in the order of increasing $i$,
shows that each of these loops must contract in turn to a point.

We thus conclude that for every $j$ such that the Landau matrix element
$$
L_{ij} \equiv \half \partial f_j/\partial \Omega_i,
$$
is non-zero for some $i$, $\alpha_j=0$.
But then the lemma represented by Eq. (5)
entails that one can distort the $\Omega_i$ contours in such a way as to
move simultaneously into the upper-half plane of each of the residue-factor
denominators $f_j$ and each of the photon-propagator denominators
$(\Omega_j)^2$. The only remaining singularities are the end-point
singularities at $r_i=0$ and $r_i=1$, and the three Feynman denominators
associated with the three $*$ lines of the $*$ graph $g$: for every other
singularity surface $f_j=0$ there is some $\Omega_i$ such that
$L_{ij} \neq 0$  for the corresponding $j$ and $i$. The consequences of
the three $*$ line singularities in conjunction with the end-point
singularities in the radial variables $r_i$ are dealt with in
papers I and III.
\newpage

\noindent{\bf Appendix A.
Proof of the triviality of the contribution from the factor $\delta (
\Omega_j \widetilde{\Omega}_j -1)$
to the Landau loop equations.}

\vskip 9pt
In discussing the singularities of the meromorphic parts in $\S 8$
we made full use of the fact that the row in the Landau matrix
corresponding to $\Omega_j\widetilde{\Omega}_j -1$ reduces to zero under
the closed loop conditions for $\Omega_j$-column, the $r_j$-column and
the
$r_{j+1}$-column.
We give here a proof of this fact.

In view of the definition of the integral, the functions $f_i$ other
than the various  $\Omega^2_{j}, \Omega_j\widetilde\Omega_{j}-1$ \ and
$r_{j}$
 have the
following form (A.1) or (A.2), where $\epsilon_m,$ \ and $ \epsilon'_t $
are each either $0$ or
$+1$
or $-1$:
$$
f_i = (p_\ell + \sum \epsilon_m r_1 ... r_m \Omega_m)^2 -m^2\eqno(A.1)
$$
$$
\eqalignno{
f_i &= 2(p_\ell + \sum \epsilon_m r_1 ... r_m \Omega_m) ( \Omega_s + \sum
\epsilon'_t r_{s+1} ... r_t\Omega_t)\cr
&+ r_1 ... r_s (\Omega_s + \sum \epsilon '_t r_{s+1} ... r_t \Omega_t)
^2.&(A.2)\cr}
$$
Let $H_j$ denote the first-order differential operator given by
$ \Omega_j {\partial\over \partial\Omega_j} - r_j {\partial\over
\partial r_j} + r_{j+1} {\partial\over \partial r_{j+1}}$.
Then the following equations hold:
$$
\eqalignno{
&H_j (p_\ell + \sum \epsilon_m r_1 ... r_m \Omega_m) =0\ \  \hbox{ for any }
\ j \ \hbox{and } \ell.
&(A.3)\cr
&H_j (\Omega_s + \sum \epsilon '_t r_{s+1} ... r_t \Omega_t)&(A.4)\cr
&=\Bigg\{ \matrix{ \Omega_s + \sum \epsilon'_{t}r_{s+1} ... r_t \Omega_t
&\hbox{if} \ j=s\cr
       0 &\hbox{if} \ j\neq s},\cr
&H_j (r_1 ... r_s (\Omega_s +  \sum \epsilon'_t r_{s+1}  ...
r_t\Omega_t)^2)&(A.5)\cr
&=\Bigg\{ \matrix{ r_1 ... r_s (\Omega_s + \sum \epsilon'_t r_{s+1} ...
r_t\Omega_t)^2 &\hbox{if} \ j=s\cr
0 &\hbox{if} \ j\neq s}\hbox{\hskip .5in}.\cr}
$$
Hence $H_j$ annihilates each $f_i$ of the form (A.1) and each $f_i$ of
the form (A.2) with $s\neq j$, and it reproduces each $f_i$ of the form (
B.2) with $s=j$.

Since the $\Omega_j$-column etc. in the Landau matrix is given by
$\partial f_i /\partial\Omega_j$ etc., this property of the operator
$H_j$ entails, under the $\Omega_j, r_j$, and $r_{j+1}$ closed-loop
conditions, that
$$
\eqalignno{
0&=\Omega_j (\sum_i \alpha_i {\partial f_i\over \partial \Omega_j}) -
r_j (\sum_i \alpha_i {\partial f_i\over \partial r_j}) + r_{j+1} (\sum_i
\alpha_i {\partial f_i\over \partial r_{j+1} })
\cr
& = \sum_i \alpha_i H_j f_i\cr
&= \sum_{i\epsilon I(j)} \alpha_i f_i + 2a_j \Omega^2_j + 2
\beta_j\Omega_j\widetilde{\Omega}_j - \gamma_j r_j + \gamma_{j+1}
r_{j+1},\cr}
$$
where $I(j)$ denotes the set of indices $i$ such that $f_i$ is of the
form (A.2) with $s = j$, and $a_j, \beta_j$ and $\gamma_j$ denote the
Landau parameters associated with
$\Omega^2_j,\ \Omega_j\widetilde{\Omega}_j -1$, and $r_j$, respectively.
It follows from (4a) that all terms except for
 $\beta_j \Omega_j\widetilde{\Omega}_j = \beta_j$
on the right-hand side of (A.6) vanish.
This entails the required fact, namely that the row corresponding to
$\Omega_j\widetilde{\Omega}_j -1$ must have coefficient $\beta_j=0$ and
hence give no net contribution to the Landau loop equations.

\newpage

\noindent{\bf Appendix B.
The Landau diagram corresponding to a term in the pole-decomposition
expansion.}
\vskip 9pt

To confirm the geometric representation of the Landau equations
described in connection with Eq.(6) recall first that the
pole-residue denominators corresponding to non $*$ charged lines are
$$
f_j = \sigma_{js} (\Sigma^2_j - \Sigma^2_s) + i0,\eqno(B.1)
$$
where the sign $\sigma_{js}$ is defined below (3).
For each side $s\epsilon \{1,2,3\} $
one may verify immediately that the contribution from the side $s$ of the
triangle of {\it direct} lines $V_s$ is just the contribution to the
$p$ loop equation arising from the charged-particle line segments that lie
on side s of the original graph.

For the photon loop $\ell$ there is first a contribution $\alpha_\ell
k_\ell$, and then the contributions corresponding to charge-particle
line segments along which the loop flows.
There are contributions of this latter kind only from segments
corresponding to those (one or two) sides $s$ of the triangle along
which the loop runs, and we can consider separately the contributions
from each of those sides $s$.

There are three cases:

Case 1.  The photon loop $\ell$ in the Feynman graph runs along the segment
$j \epsilon J(s)$ but does {\it not} run
along the $*$ segment lying on side $s$.
In this case the contribution to the $\ell$ loop equation proportional to
$\alpha_j$ is
$$
\eqalignno{
\alpha_j \half {\partial f_j\over \partial k_\ell} &=
\sigma_{js} \alpha_j \half {\partial\over \partial k_\ell} (\Sigma^2_j
- \Sigma^2_s)
\cr
&= \beta_{js} \Sigma_j.&(B.2)\cr}
$$

Case (2a). The loop $\ell$ of the Feynman graph flows along the $*$
segment of side $s$, but does {\it not } flow along the non $*$ segment $j$
lying on side $s$.
Then the contribution to the $\ell$ loop equation proportional to $\alpha_j$
is
$$
\eqalignno{
\alpha_j \half {\partial f_j\over \partial k_\ell} &= \sigma_{js} \alpha_j
\half
{\partial\over \partial k_\ell} (\Sigma^2_j - \Sigma^2_s)\cr
&= \beta_{js} ( - \Sigma_s).&(B.3)\cr}
$$

Case (2b).  The loop $\ell$ of the Feynman graph flows along the $*$ segment of
side $s$ of the graph and
also along the non $*$ segment $j \epsilon J(s)$.
Then the contribution to the $\ell$ loop equation proportional to
$\alpha_j$ is
$$
\eqalignno{
\alpha_j \half {\partial f_j\over \partial k_\ell} &= \beta_{js} \alpha_j \half
{\partial\over \partial k_\ell} (\Sigma^2_j - \Sigma^2_s)\cr
&= \beta_{js} (\Sigma_j - \Sigma_s) . &(B.4)\cr}
$$

Notice that, according to (B.2), (B.3), and (B.4),there is, for each
$j \epsilon J(s)$,  a contribution
$\beta_{js} \Sigma_j$ to the photon loop equation $\ell$ if and only if the
loop $\ell$ in the graph passes along the segment $j$. There is also,
for each $j \epsilon J(s)$, a contribution $-\beta_{js} \Sigma_s$ if and only
if this loop passes along the star line $s$ in the graph.
There is also a contribution $\alpha_s \Sigma_s $ if and only if
this loop passes along the star line $s$ of the graph.
These results are summarized by the rules (6).
\newpage

\newpage
\noindent{\bf References}
\begin{enumerate}
\item T. Kawai and H.P. Stapp, {\it Quantum Electrodynamics at Large
Distances I: Extracting the Correspondence-Principle Part.}
Lawrence Berkeley Laboratory Report LBL 35971 (1994), Submitted to Phys. Rev.;
See also H.P. Stapp, Phys. Rev. {\bf D28},1386 (1983)
\item H.P. Stapp, in Structural Analysis of Collision Amplitudes, ed.
 R. Balian and
D. Iagolnitzer, North-Holland, New York, 1976, p.200 (Pham's Theorem).
\item L.D. Landau, in Proc. Kiev Conference on High-Energy Physics (1959);
Nuclear Physics {\bf 13} (1959) 181; N. Nakanishi, Prog. Theor. Phys. {\bf 22}
(1959) 128.
\item R.J. Eden, P.V. Landshoff, D.I. Olive, and J.C. Polkinghorne,
The Analytic S-matrix, Cambridge University Press, p.57 (1966);
\item T. Kawai and H.P. Stapp, {\it in} Algebraic Analysis, eds. M. Kashiwara
and T. Kawai, Acad. Press (1988) Vol. I
\item T. Kawai and H.P. Stapp {\it Publ.} RIMS {\it Kyoto Univ.} {\bf 12}
Suppl.155 (1977) Theorem 2.1.1
\item J. Coster and H.P. Stapp, J. Math Physics {\bf 11} (1970) 2743 (p. 2758)
\item C. Chandler and H.P. Stapp, J. Math. Phys. {\bf 10} (1969) 826.
\item D. Iagolnitzer and H. P. Stapp, Commun. Math. Phys. {\bf 14} (1969) 15.
\item H.P. Stapp, in Structural Analysis of Collision Amplitudes, ed.
R. Balian and
D. Iagolnitzer, North-Holland, New York, (1976).
\item D. Iagolnitzer, The S-matrix, North-Holland, (1978); Comm. Math. Phys.
{\bf 41}, 39 (1975); Comm. Math. Phys. {\bf 63}, 49 (1978)

\end{enumerate}

\end{document}